\documentclass[runningheads]{llncs}
\usepackage[T1]{fontenc}
\usepackage{graphicx}
\begin{document}
\title{Investigating Gender Bias in Lymph-node Segmentation with Anatomical Priors}
%

\author{Ricardo Coimbra Brioso\inst{1} \and  Damiano Dei \inst{2,3} \and Nicola Lambri\inst{2,3} \and Pietro Mancosu\inst{2,3} \and Marta Scorsetti\inst{2,3} \and Daniele Loiacono\inst{1}}

\institute{Department of Electronics, Information and Bioengineering, Politecnico di Milano, Milan, Italy \and Department of Biomedical Sciences, Humanitas University, Milan, Italy \and Radiotherapy and Radiosurgery, IRCCS Humanitas Research Hospital, Rozzano, Italy}

\maketitle              
\abstract{Radiotherapy requires precise segmentation of organs at risk (OARs) and of the Clinical Target Volume (CTV) to maximize treatment efficacy and minimize toxicity. While deep learning (DL) has significantly advanced automatic contouring, complex targets like CTVs remain challenging. This study explores the use of simpler, well-segmented structures (e.g., OARs) as Anatomical Prior (AP) information to improve CTV segmentation. We investigate gender bias in segmentation models and the mitigation effect of the prior information. Findings indicate that incorporating prior knowledge with the discussed strategies enhances segmentation quality in female patients and reduces gender bias, particularly in the abdomen region. This research provides a comparative analysis of new encoding strategies and highlights the potential of using AP to achieve fairer segmentation outcomes.}


\section{Introduction}
For several types of cancer, radiotherapy is the most effective treatment, and it is used in more than 50\% of cancer patients as the main treatment, concurrent, or perioperative multimodal treatments~\cite{Baskar2012CancerAR}.
Radiotherapy uses ionizing radiation to kill cancer cells and shrink tumors.
To avoid toxicity and side effects, the radiation dose must be delivered to the tumor while sparing the surrounding healthy tissues.
Accordingly, radiation oncologists (ROs) define the OARs and the CTV which is the volume that corresponds to the tissue to be irradiated and includes a small margin that accounts for body changes and movement.
The ROs dedicate several hours to this contouring in the full-body Computed Tomography (CT) and the automation of these segmentations would speed up this process.

Advancements in DL proved to be a very effective and consistent tool for automating the image contouring process in radiotherapy~\cite{huynh2020artificial}. Healthcare centers and hospitals are beginning to use tools embedded with DL models to automate contouring.
This is possible due to research in DL, hardware improvements, and, data availability.
Medical imaging tasks are characterized by having a smaller amount of data in comparison to other computer vision tasks. This coupled with the complexity of some targets, such as CTV in Total Marrow and Lymph node Irradiation (TMLI) ~\cite{mancosu2020}, increases the difficulty of improving automatic segmentation in this particular task.

Segmentation models can perform better based on architectural changes in the network, this has motivated iterative improvements in models' performance. In this article, it is studied an alternative path for improving the segmentation quality. Available segmented structures that are simpler than the CTV (e.g.: OARs) are available in more datasets and can be obtained with open-source models \cite{Wasserthal2022}. Using OARs and other structures as prior information to the segmentation model is still unexplored. This work provides the first comparative analysis of strategies to encode prior knowledge.

While assessing the models, gender disparity is found in the segmentation performance. Gender bias has always been present in many DL tasks but research in this field is limited despite its importance. Approaches to mitigate gender bias in segmentation models are even less common \cite{Puyol2021_miccai}. To accurately measure fairness, different metrics are applied and the segmentation results are observed in different regions of the patient.
The findings of this work move us towards exploring Anatomical Priors (AP) to mitigate gender bias. Using already available data that contains gender-specific information improves the model's performance in the abdomen regions. Even structures that exist in both genders can have information about the gender, for example, the size and ratio of the structure when compared to other structures can vary between female and male.
The main contributions of this work are: (i) exploring and comparing new encoding strategies for AP in the segmentation model's training.
(ii) analyzing gender bias and the effects of AP in mitigating it in the different regions of the body.

\section{Related Work}
Radiotherapy has automatic segmentation solutions for OARs and CTV, although CTV remains rarer to see segmented. For OARs and other structures, we can see works for different regions: the head and neck~\cite{Liu2020}, thorax~\cite{Yang2020}, abdomen~\cite{Tong2020}, and pelvis~\cite{Ma2022}.

Cervical cancer CTV segmentation in \cite{Liu2020_2} is obtained using DpnUNet, a similar architecture to the U-Net \cite{unet}, it combines Residual and Dense blocks in the encoder, enhancing the ability to recover and refine abstract features. It also adds three adjacent slices as three new channels in the input of the network, making it a 2.5D architecture.
In \cite{Song2020}, segmentation models for rectal and cervical cancer CTV are developed using U-Net-based architectures or DeeplabV3+ \cite{deeplabv3p}. The work of \cite{Song2020} concludes that time is saved by using the CTV automatic contouring, even when the latter needs corrections. In \cite{Ma2022}, an automatic CTV contouring is compared with an RO's contouring and observes that automatic contours are on par with manual contouring and save RO's time.

The CTV is a complex agglomerate of structures that is hard to segment, in some works it is shown that adding anatomical prior information about the segmentation target and incorporating it in the loss function can improve its performance and anatomical plausibility \cite{Brioso2023_2}.

In \cite{Lian2023}, several single-organ models are used to learn anatomical invariance across
different subjects and datasets and using this information as priors for other segmentation models, thus, improving performance. To focus the attention of the network in the pancreatic area, in \cite{Shen2024}, a mask indicating the region of the pancreas is added as a prior.

Following the CTV segmentation experiments with AP, it was analyzed that this addition mitigated the gender bias present in the segmentation performance. Analyzing biases in deep learning models is an important topic that has only gained visibility in recent years \cite{Buolamwini2018}.
In medical imaging segmentation, only a handful of articles investigate model biases. In \cite{Puyol2021_miccai}, several training approaches were developed that helped to mitigate racial bias.
For example, a stratified batch to ensure a balance between racial groups in each iteration of the training or an additional DL classification network that classified the gender before performing the training segmentation.

The work \cite{lee2023investigation} compared gender and race bias in three DL segmentation architectures and a transformer architecture in cardiac MR segmentation. The models' biases differ for each gender or race and respond differently to the percentage of minority group's data present in the training set of the model.

Anatomical priors and gender analysis works for segmentation models are scarce but very important due to their applicability and consequences in the real world.
For the first time, diverse ways to encode AP information are compared, while analyzing its effects on gender bias of segmentation models.

\section{Methods}

\subsection{Data}

The data used contains 45 full-body CT 3D volumes, 25 male and 19 female patients.  A free-breathing, non-contrast CT scan with a 5-mm slice thickness was acquired for each patient using a BigBore CT system (Philips Healthcare, Best, Netherlands) \cite{mancosu2021} and each axial slice of the CT has a resolution of 512x512 pixels. The patients are candidates to undergo radiotherapy treatment and for every CT volume, several structures were delineated by a RO, including the Clinical Target Volume (CTV) that corresponds to the goal target of this segmentation task. For two steps of the experimental design, structures generated with the help of TotalSegmentator were used. Firstly, several structures from TotalSegmentator were used as AP to train the segmentation models. Secondly, the vertebra T1, the stomach, and, vertebra L4. were used to separate each patient into four regions: the Head and Neck (HN), the thoracic (THX), the abdomen (ABDM), and the pelvic region (PELV). The separation of the regions is used to evaluate the performance of the CTV segmentation in more detail in each region.

\subsection{Anatomical Prior Strategies}

Several experiments were made by inputting anatomical prior information (segmentation of other structures or OARs). The APs were inputted in different ways:
Multiple Intensity Z-Score (MI-Z): The structures (spleen, liver, eyes, kidneys, femurs, stomach, heart) present in the ground truth (GT) of the dataset were added as an additional channel to the input after encoding each structure with a different value of intensity (from 0 to 255) with Z-Score normalization on the APs.
Equal Intensity Z-Score (EQ-Z): Similar to the previous inputting but the value of intensity is the same for every structure (intensity of 255).
Cropped CT Z-Score (Crop-Z): Two additional channels were used in this inputting strategy, both of them included the original intensities of the CT image in the image's position of the additional structures, while the image's external part to these structures had an intensity value of 0 in one channel and 255 in the other channel.
Multiple Intensity (MI): Identical to the MI-Z inputting but without Z-Score normalization.
Multiple Intensity with TotalSegmentator (MI-TS): The structures used in this inputting technique were not from the GT but from TotalSegmentator and no normalization was applied. The structures used in this case were: humeri, scapulae, clavicles, femurs, hips, sacrum, spleen, liver, stomach, urinary bladder, pancreas, kidneys, and, iliopsoas muscles. The structures were encoded in the same way as the first strategy, each with its intensity and without Z-Score normalization on the APs.

Providing contextual information from adjacent organs could improve segmentation performance and reduce sex-based bias.

\subsection{Training and Evaluation}

We applied the nnU-Net \cite{nnUnet} framework, an adapting algorithm that analyzes the dataset’s characteristics, such as resolution and pixel spacing. This information is used for pre-processing and tuning the training parameters of the nnU-Net.

The evaluation metrics used were the Dice Score (DSC) and Hausdorff Distance (HD).

The $DSC = \frac{2 |X \cap Y| }{|X| + |Y|}$ measures the overlap between the GT and the predicted mask, with $X$ representing the set of positive pixels in the GT and $Y$ representing the set of positive pixels in the prediction.

The $HD=max\left\{\max_{x\in S_X} d(x,S_Y), \max_{y\in S_Y} d(y,S_X) \right\}$ measures the maximum distance of all the nearest distances between the surfaces of the two sets $X$ and $Y$, denoted as $S_X$ and $S_Y$. To mitigate the impact of outliers on HD values, we employed HD95, which excludes the top 5\% highest HD values.

An evaluation focused on different regions of the patient was conducted to localize the model's biases.
To evaluate the performance of the CTV segmentation model in different regions of the patient, structures segmented by TotalSegmentator were used to locate the boundaries of four regions of the patient's body: HN, THX, ABDM, PELV. Dividing the patient analysis is useful to see if the bias is related to a specific part of the patient that is harder to segment.
Several metrics were developed to understand the underlying patterns and location of the gender bias:

\begin{itemize}
    \item \textbf{Average Gender Difference (AGD)}: The average difference in Dice scores between male and female patients.
    \item \textbf{Median Gender Difference (MGD)}: Measures the median difference in Dice scores between male and female patients.
    \item \textbf{Quartile Difference (QD)}: Calculated as the maximum of the difference between the third quartile of male DSC and the first quartile of female DSC, and the difference between the third quartile of female DSC and the first quartile of male DSC.
\end{itemize}
Due to the quantity of available data, a statistical analysis is not presented.

\section{Results and Discussion}
In Table \ref{resultsBias}, we can see the whole-body performance of the different AP models and their respective gender biases. All models have similar median DSC values, ranging from 82.98\% to 83.46\%, indicating comparable segmentation performance across models. The HD95 median across all models is around 4.8mm to 4.9mm, suggesting minimal changes on the surface of the contouring with AP. The MI-TS model has the highest median DSC at 83.46\%, suggesting a slight edge in segmentation performance.

The Base model shows a median DSC difference of 5.25\% between females (79.61\%) and males (84.86\%), indicating gender bias. There is a gender median DSC disparity in every model, varying from 3.82\% to 5.33\%. MI-Z improved DSC performance for females (80.58\%) compared to the base model, and a slightly lower performance for males (84.40\%), having one of the smallest median DSC differences between males and females across all models. This indicates a reduction in gender disparity. The MI model improved median DSC while maintaining a high male median DSC. Models MI-Z, EI-Z, Crop-Z, MI, and MI-TS show a trend toward reducing gender bias, improving the DSC for female patients without significantly compromising the performance of male patients.

\begin{table}[]
\centering

\caption{Different Multiple-Input Models and their Gender Bias}
\label{resultsBias}
\begin{tabular}{ccccc}
\textbf{Model } & \textbf{DSC Med.} & \textbf{HD95 Med.} & \begin{tabular}[c]{@{}c@{}}\textbf{DSC}\\ \textbf{F-Med.}\end{tabular} & \begin{tabular}[c]{@{}c@{}}\textbf{DSC}\\ \textbf{M-Med.}\end{tabular} \\ \hline
Base   & 83.40\%  & 4.83      & 79.61\%                                              & 84.86\%                                              \\
MI-Z   & 83.36\%  & 4.76      & 80.58\%                                              & 84.40\%                                              \\
EI-Z   & 83.26\%  & 4.83      & 80.35\%                                              & 84.32\%                                              \\
Crop-Z & 83.14\%  & 4.76      & 80.50\%                                              & 84.59\%                                              \\
MI     & 82.98\%  & 4.91      & 80.46\%                                              & 85.33\%                                              \\
MI-TS  & 83.46\%  & 4.69      & 80.24\%                                              & 84.20\%                                             
\end{tabular}
\end{table}

Table \ref{Table:resultsBias_regions} shows that the Base model has a high median DSC across all regions, with values ranging from 82.63\% (PELV) to 85.31\% (HN).
There is a notable gender disparity, with males having a higher median DSC than females in all regions. In the HN region, males achieve a median DSC of 85.90\% compared to 82.90\% for females.
The MI-Z model improves the median DSC in the ABDM region to 86.00\%, indicating an enhancement in segmentation performance in this region while showing a decrease in the HN (84.84\%) and THX (82.97\%). 

The EI-Z model maintains consistent performance across regions, with slight improvements in the THX (83.12\%) and ABDM (85.76\%) regions. The Crop-Z had the greatest improvement for the HN region in general and for male patients while the MI-TS improved the HN performance for female patients ($\uparrow 0.3\%$).

The MI-TS model shows an improvement in the THX region with a median DSC of 83.10\%.
Performance in other regions is slightly lower compared to the Base and MI models.
Gender disparities are observed, with males consistently achieving higher median DSCs.

\begin{figure}
	\centering
		\includegraphics[scale=0.75]{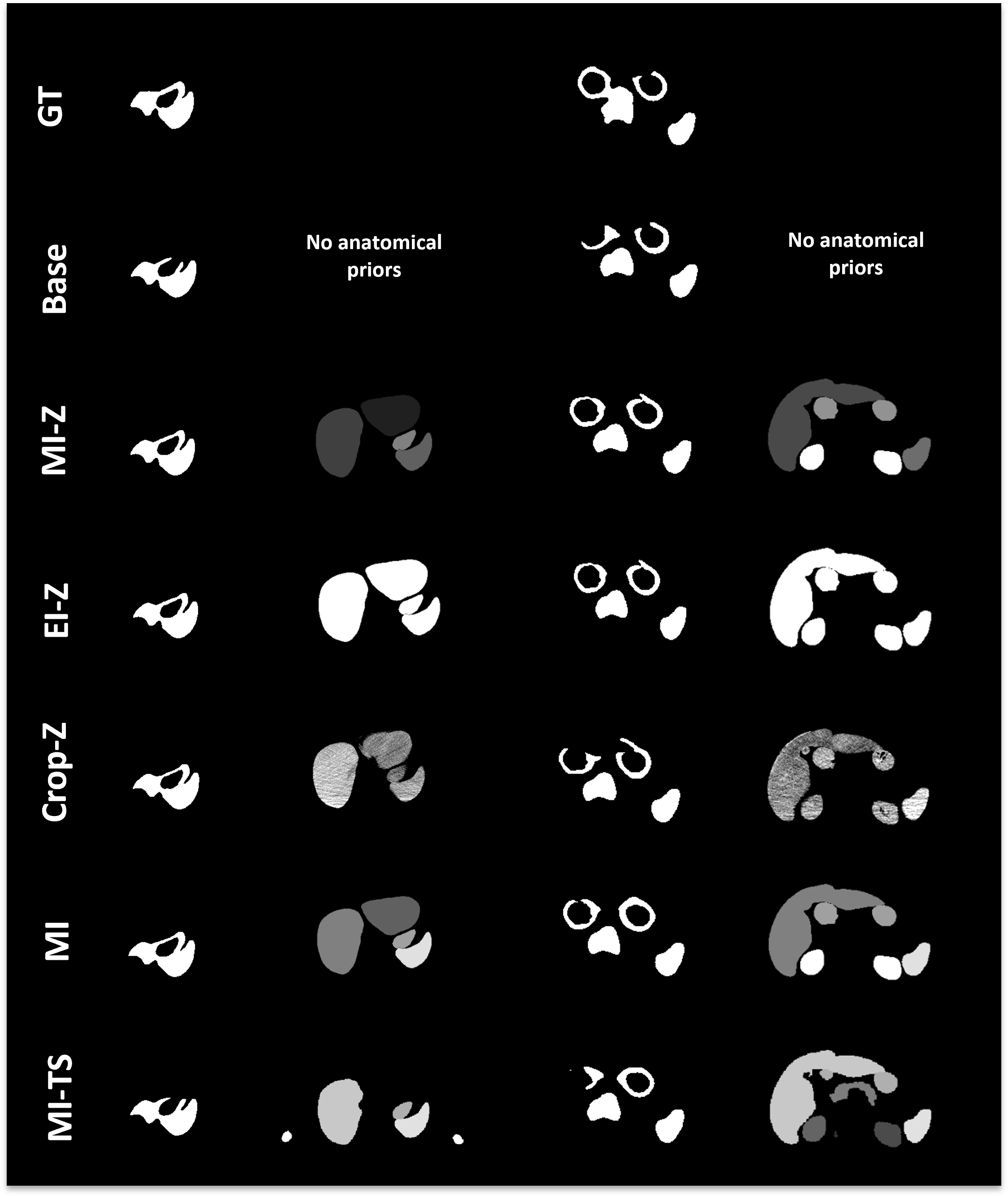}
	\caption{Examples of improvements in the CTV segmentation predictions due to anatomical prior addition in the ABDM region.}
	\label{anatomicalPriorVisual}
\end{figure}

The region that improved the most was the ABDM region, where all the AP models improved DSC for female patients. 
This outcome is related to the high number of simultaneous AP structures in this region, as seen in Figure \ref{anatomicalPriorVisual}. In Figure \ref{anatomicalPriorVisual}, the first two columns correspond to an example where having APs of the heart, the liver, and the stomach, improves the segmentation of the lymph nodes around the stomach (lymph nodes are part of the CTV). The second example of another patient (located in the last two columns) shows that the AP of the stomach improved the delineation of the surrounding lymph nodes once more.

\begin{table}[]
\centering

\caption{Different Multiple-Input Models and their Gender Bias in different regions.}
\label{Table:resultsBias_regions}
\begin{tabular}{cccccc}
\textbf{Model} & \textbf{Med. DSC $\uparrow$} & \textbf{HN}      & \textbf{THX}     & \textbf{ABDM}    & \textbf{PELV}    \\ \hline
Base           & Total             & 85.31\%          & \textbf{83.58\%} & 84.34\%          & \textbf{82.63\%} \\
               & Female            & 82.90\%          & 79.90\%          & 82.50\%          & \textbf{81.20\%} \\
               & Male              & 85.90\%          & 85.20\%          & 86.30\%          & 83.10\%          \\ \hline
MI-Z           & Total             & 84.84\%          & 82.97\%          & \textbf{86.00\%} & 82.36\%          \\
               & Female            & 79.50\%          & \textbf{80.40\%} & \textbf{85.80\%} & 80.80\%          \\
               & Male              & 86.50\%          & 85.80\%          & 86.30\%          & 83.00\%          \\ \hline
EI-Z           & Total             & 84.63\%          & 83.12\%          & 85.76\%          & 82.30\%          \\
               & Female            & 79.90\%          & 79.90\%          & 85.70\%          & 80.90\%          \\
               & Male              & 86.10\%          & 85.40\%          & 86.30\%          & 83.20\%          \\ \hline
Crop-Z         & Total             & 85.35\%          & 82.86\%          & 85.58\%          & 82.34\%          \\
               & Female            & 79.40\%          & 80.20\%          & 85.00\%          & 80.50\%          \\
               & Male              & 86.30\%          & 85.50\%          & 85.90\%          & 82.90\%          \\ \hline
MI             & Total             & 85.92\%          & 82.59\%          & 84.62\%          & 82.50\%          \\
               & Female            & 83.10\%          & 80.10\%          & 84.30\%          & 80.80\%          \\
               & Male              & \textbf{86.90\%} & 85.80\%          & 84.90\%          & 82.80\%          \\ \hline
MI-TS          & Total             & 84.97\%          & 83.10\%          & 84.50\%          & 82.05\%          \\
               & Female            & \textbf{83.20\%} & 80.00\%          & 83.20\%          & 81.00\%          \\
               & Male              & 86.30\%          & 85.20\%          & 86.40\%          & 82.50\%         
\end{tabular}
\end{table}

Table \ref{resultsBIAS_other_metrics_regions} shows three fairness metrics: AGD, MGD, and, QD for the AP models.

The MI-TS model shows the lowest AGD and MGD for HN, THX, and PELV. The regions with higher disparity are the HN, THX, and ABDM. The PELV region contains the majority of gender-specific structures but has the lowest disparity, this could indicate that the CTV in this region is easier to segment due to its relatively lesser area in this region and that the APs used were insufficient to help the model improve its performance.
Despite some models reducing AGD and MGD, the regions remain with high QD favoring males, reflecting bias.
The insights provided by fairness metrics are crucial for selecting and improving segmentation models and ensuring fairness and similar accuracy across different patient demographics and real-life applications.

\begin{table}[]
\centering
\caption{Different Multiple-Input Models and their Gender Bias in different regions for the developed fairness metrics.}
\label{resultsBIAS_other_metrics_regions}
\begin{tabular}{ccccc}
\textbf{Model} & \textbf{HN}      & \textbf{THX}     & \textbf{ABDM}   & \textbf{PELV}   \\ \hline
\multicolumn{5}{c}{\textbf{AGD $\downarrow$ }}                                                          \\
Base            & 3.30\%           & 2.70\%           & \textbf{1.80\%} & 1.30\%          \\
MI-Z            & 3.70\%           & 3.00\%           & 2.30\%          & 1.70\%          \\
EI-Z            & 3.70\%           & 2.70\%           & 2.00\%          & 1.30\%          \\
Crop-Z          & 3.70\%           & 2.80\%           & 2.50\%          & 1.20\%          \\
MI              & \textbf{3.00\%}  & 2.60\%           & 2.40\%          & 1.40\%          \\
MI-TS           & \textbf{3.00\%}  & \textbf{2.30\%}  & 2.40\%          & \textbf{1.10\%} \\ \hline
\multicolumn{5}{c}{\textbf{MGD $\downarrow$}}                                                          \\
Base            & \textbf{3.00\%}  & 5.30\%           & 3.80\%          & 1.90\%          \\
MI-Z            & 7.00\%           & 5.50\%           & \textbf{0.50\%} & 2.10\%          \\
EI-Z            & 6.20\%           & 5.50\%           & 0.60\%          & 2.30\%          \\
Crop-Z          & 6.90\%           & 5.40\%           & 1.00\%          & 2.30\%          \\
MI              & 3.80\%           & 5.70\%           & 0.60\%          & 2.00\%          \\
MI-TS           & \textbf{3.20\%}  & \textbf{5.20\%}  & 3.20\%          & \textbf{1.50\%} \\ \hline
\multicolumn{5}{c}{\textbf{QD $\downarrow$}}                                                           \\
Base            & 12.90\%          & 10.60\%          & \textbf{9.90\%} & \textbf{5.90\%} \\
MI-Z            & 13.40\%          & 12.50\%          & 11.80\%         & \textbf{5.90\%} \\
EI-Z            & \textbf{12.10\%}          & 12.40\%          & 11.70\%         & \textbf{5.90\%} \\
Crop-Z          & 12.90\%          & 12.30\%          & 10.60\%         & \textbf{5.90\%} \\
MI              & 12.50\%          & 11.90\%          & 10.90\%         & 6.50\%          \\
MI-TS           & \textbf{12.10\%} & \textbf{10.20\%} & 10.20\%         & 7.00\%         
\end{tabular}
\end{table}

\section{Conclusions and Future Work}
While the base model serves as a solid baseline, AP models like MI-Z, and MI-TS tend to reduce gender disparity in segmentation performance, especially in the ABDM region, indicating that these models are capable of handling the prior information to improve and mitigate gender-related anatomical differences.
This study is limited by our small patient number and future work should be focused on testing different inputting prior strategies and their bias in larger datasets and architectures. Choosing the appropriate structures to add contextual information and mitigate should include the aid of clinicians. Studying other commonly recurring biases such as racial and size of the patient bias should be thought about as well.

\bibliographystyle{splncs04}
\bibliography{mybibliography}

\end{document}